\documentclass{ws-ijmpb}

\begin{document}

\markboth{Patrick Grosfils and Jean Pierre Boon}
{Viscous Fingering in miscible, immiscible and reactive fluids}

\catchline{}{}{}

\title{VISCOUS FINGERING IN MISCIBLE,\\ 
       IMMISCIBLE AND REACTIVE FLUIDS}

\author{\footnotesize PATRICK GROSFILS \footnote{
E-mail: {\tt pgrosfi@ulb.ac.be}}}

\address{Center for Nonlinear Phenomena and Complex Systems\\
Universit\'{e} Libre de Bruxelles, 1050 - Bruxelles, Belgium
}

\author{JEAN PIERRE BOON \footnote{
E-mail: {\tt jpboon@ulb.ac.be};
http://poseidon.ulb.ac.be}}

\address{Center for Nonlinear Phenomena and Complex Systems\\
Universit\'{e} Libre de Bruxelles, 1050 - Bruxelles, Belgium
}

\maketitle

\pub{Received (received date)}{Revised (revised date)}

\begin{abstract}

With the Lattice Boltzmann method (using the BGK approximation)
we investigate the dynamics of  Hele-Shaw flow under
conditions corresponding to various experimental systems.
We discuss the onset of the instability (dispersion relation),
the static properties (characterization of the 
interface) and the dynamic properties (growth of the mixing zone) 
of simulated Hele-Shaw systems. We examine the role of 
reactive processes (between the two fluids) and we show that 
they have a sharpening effect on the interface similar to 
the effect of surface tension.  
\end{abstract}

\bigskip

Viscous fingering occurs in the interfacial region between 
two fluids with different viscosities when a highly viscous 
fluid confined between two plates with a narrow gap (Hele-Shaw
geometry) is displaced by a fluid with relatively low viscosity.
Most theoretical and numerical studies on Hele-Shaw systems
start with Darcy's law~\cite{darcy,homsy} assuming its validity under 
the simulation conditions, and thereby precluding discrimination
between Navier-Stokes behavior or Darcy behavior of the system.
With the Lattice Boltzmann (LB) approach \cite{succi} the 
discrimination problem can be addressed because the LB method  
starts with a kinetic theoretical analysis where the 
macroscopic description is not pre-established.

While the LB method has now become a standard approach to
investigate fingering processes, it has mostly been applied
to the formation and growth of a single finger in a channel
\cite{yoemans}. Fingering in large systems have been studied 
and analyzed with Lattice Gas Automata constructed to simulate 
porous media \cite{lutsko,hayot}. Here we consider fingering 
in spatially extended Hele-Shaw systems. 
We investigate the effects of surface tension and reactivity 
between the two fluids, and we demonstrate their role as 
determinant factors in the dynamics of the moving interface.
The present paper gives a presentation of the essentials of
the method and describes the main results of the work which
will be discussed in detail elsewhere.   


We use the LB equation with the Bhatnagar-Gross-Krook collision 
operator (LBGK method; see e.g. \cite{succi}) in two-dimensional 
geometry emulating 3-D flows by introducing a Hele-Shaw drag term 
thereby simulating a system with a virtual cell gap in the third 
dimension. Using standard notation, the starting LBGK equation 
for multicomponent systems reads
\begin{equation}
f_i^\sigma(r+c_i,t+1)\,-\,f_i^\sigma(r, t)\,=\,-\frac{1}{\tau^\sigma}\,
(f_i^\sigma\,-\,f^{^\sigma\rm eq}_i)\,,
\label{lbgk}
\end{equation}
where $\sigma$ denotes the label of the component with density
$\rho^\sigma(r)\,=\,\sum_i\,m^\sigma\,f_i^\sigma$ and velocity
${\bf u}^\sigma(r)\,=\,\frac{m^\sigma}{\rho^\sigma(r)}\,
\,\sum_if_i^\sigma {\bf c}_i$.
Taking the long wavelength limit of the LBGK equation, one recovers 
the phenomenological equations of Navier-Stokes hydrodynamics 
(see e.g. \cite{lutsko}).  

Additional features can be incorporated in the LBGK equation
to account for the physics of the problem to be considered.
Here we introduce (i) an {\it interaction potential} 
(initially used to simulate non-ideal fluids \cite{doolen})
for tunable miscibility (``surface tension''); 
(ii)~an {\it external force} for the Hele-Shaw drag term 
(a term that can also be used to incorporate tunable gravitational 
effect in density driven fingering), and 
(iii) a {\it reactive term} which quantifies 
changes in the ``chemical'' nature of the species which
are susceptible to react with each other. 

\noindent (i) The interaction potential has the form
$V(r, r')\,=\,G_{\sigma\sigma'}(r, r')\,{\rho}^{\sigma}(r)
{\rho}^{\sigma'}(r')$
with  $G_{\sigma\sigma'}(r, r')\,=\,
G_{\sigma\sigma'}(r-r')\,=\,G_{\sigma\sigma'}$ for 
$(r-~r'~)\,\equiv \,|{\bf r-r'}|\,=\,c$, and 
$G_{\sigma\sigma'}(r, r')\,=\,0$ for $|{\bf r-r'}|\,>\,c$,
$c$ being the modulus of the relative velocity between particles.
The potential enters the dynamics through the modified local velocity
\begin{equation}
\rho^\sigma(r)\,\tilde{\bf u}(r)\,=\,\rho^\sigma(r){\bf u}(r)\,+\,
\tau^\sigma\,{\bf F}^\sigma(r,t)\,,
\label{force_eq}
\end{equation} 
with
\begin{equation}
{\bf F}^\sigma(r,t)\,=\,
-{\rho}^\sigma(r)\,\,\sum_{\sigma'} G_{\sigma\sigma'}\,\sum_i\,
{\rho}^{\sigma'}(r+c_i)\,{\bf c}_i \,.
\label{force}
\end{equation}
The amplitude ($G$) of the interaction term can be tuned from 
zero (miscible fluids) to practically total immiscibility.    

\noindent (ii) The external force (here the Hele-Shaw drag term) 
is introduced by a force term which modifies the momentum of 
species $\sigma$; the new momentum reads
\begin{equation}
\rho^\sigma(r)\,\tilde{\bf u}^\sigma(r)\,=\,
\rho^\sigma(r)\,{\bf u}^\sigma (r)\,
(1\,-\,\beta_\sigma\, \tau^\sigma)\,.
\label{momentum}
\end{equation}
For the classical situation where fluid $1$ invades fluid $2$,
the tunable drag coefficient is the ratio $\beta_1 / \beta_2$.

\noindent (iii) Reactive processes are taken into account by a 
formulation similar to that used in reaction-diffusion equations 
(see e.g. \cite{boon}). Here we shall use a typical reactive term 
$\sim \kappa \,\rho_1\,\rho_2\,(\rho_1-\rho_2)$ for species $1$
(and similarly for species $2$ with $1\Longleftrightarrow 2$) 
where $\kappa $ is the kinetic constant. This form restricts the
reactivity effects to the mixing zone~\cite{dewit}. 


The LBGK model sketched above was used for 
binary fluids ($\sigma=1,2$) on the $2-D$ square lattice with 
nine velocities oriented along the lattice axes and the bisecting 
directions plus a zero speed (D2Q9 model \cite{qian}).
The simulations were performed on a $L_x \times L_y = 
1024 \times 512$ lattice with the following initial 
conditions: fluid $1$ with lowest viscosity is injected from
the upper boundary of the system filled with the highly viscous 
fluid $2$; periodic boundary conditions are imposed on the
vertical boundaries; the initial planar interface is perturbed 
with white noise to trigger the instability. As time evolves, 
the highly viscous fluid $2$ is displaced by fluid $1$. 
The color code used in Fig.1 is from dark gray to light 
gray for $\rho_2 = 0 \rightarrow 1$ so that the pictures show 
``dark'' fluid invading from above the system initially filled 
with ``light'' fluid.

Preliminary simulations were performed with a single phase fluid 
subject to a density gradient ($\nabla \rho$) imposed along 
the vertical axis ($y$), and the average velocity $v_y$ 
was measured for various values of the gradient: 
$v_y$ was found to be a linear function of the pressure gradient 
($\propto \nabla \rho$) according to Darcy's law.

The top panel in Fig.1 illustrates the case of miscible fluids
($G=0$ in Eq.(\ref{force})) where a distinguishable mixing layer
can be quantified to measure the mixing length $L_{mix}$ as a 
function of time $t$ for increasing values of the P\'eclet number
($Pe = L_x v_y /D$ where $D$ is the diffusion coefficient). 
In agreement with experimental observation and numerical simulation 
\cite{experiment}, we observe a transition from the short time 
diffusive regime where $L_{mix}$ scales as $t^{\alpha}$ with 
${\alpha} \simeq 0.5$ to ${\alpha} \simeq 1$ behavior 
characteristic of the non-linear regime. 

The middle panel of Fig.1 shows the effect of surface tension
on the finger topology (compare with upper panel). Increasing
the amplitude ($G$) of the interaction term reduces drastically
the width of the mixing layer which eventually vanishes, and 
the finger wavelength is modified correspondingly. A quantification
of this effect is given in Fig.2 where we show typical 
dispersion curves.

Reactive processes between two moving miscible fluids are expected 
to modify the fingering mechanisms as shown by phenomenological 
approaches based on a reaction-diffusion scheme \cite{dewit}, and 
experimental studies have indeed demonstrated that the interface
properties depend strongly on chemical reactions 
\cite{experim_reac}. We have implemented a reactive mechanism 
by incorporating in the LB scheme a reaction term as described
in (iii) above. The resulting effects are illustrated in the lower 
panel of Fig.1 which shows (i) a sharpening of the interfacial
zone, and (ii) a modification of the topology of the fingering
(compare with top panel) as quantified by the dispersion curves
shown in Fig.2.

\begin{figure}[htbp]
\begin{center}
\resizebox{10cm}{!}
{\includegraphics{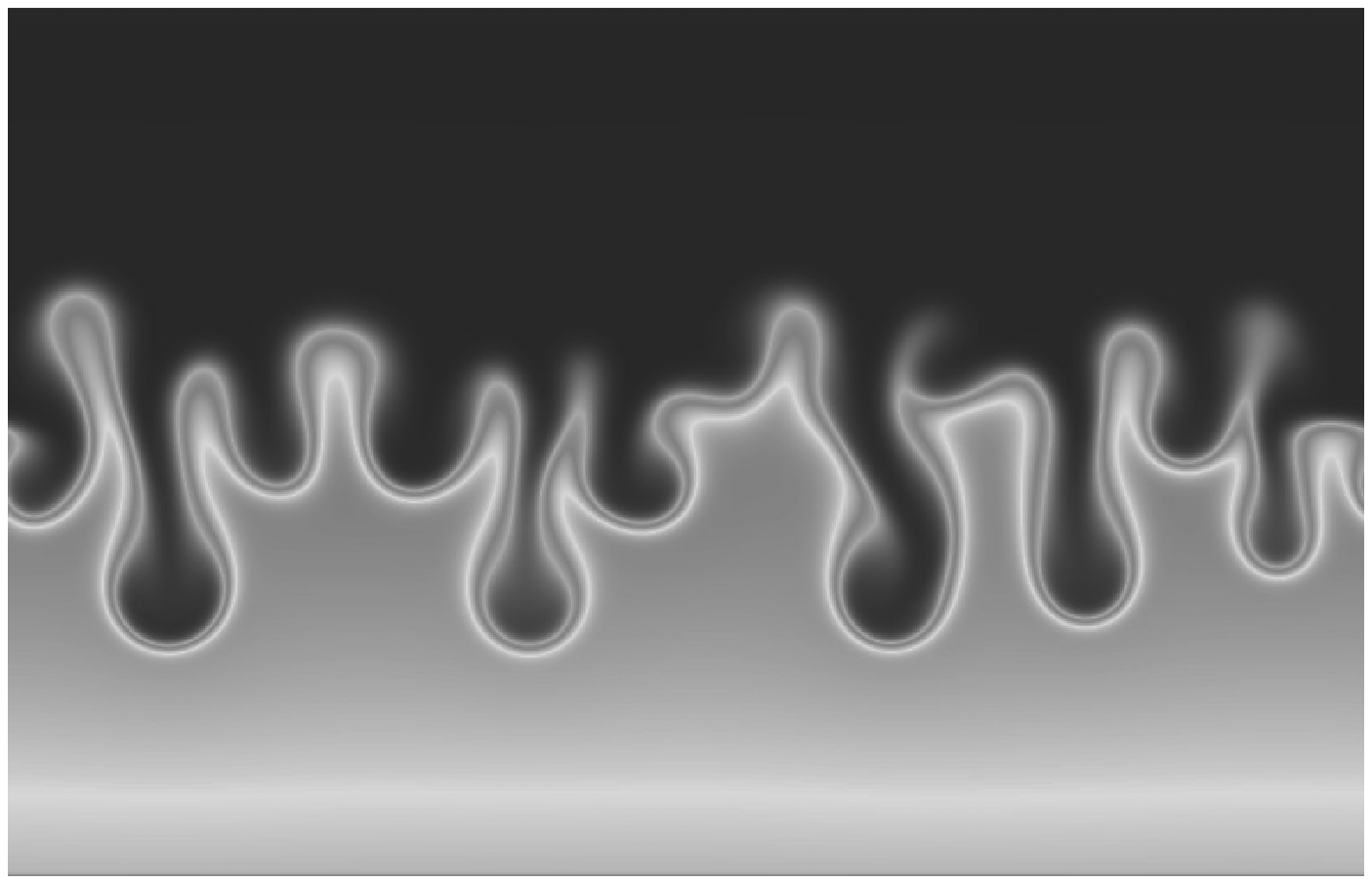}}
\resizebox{10cm}{!}
{\includegraphics{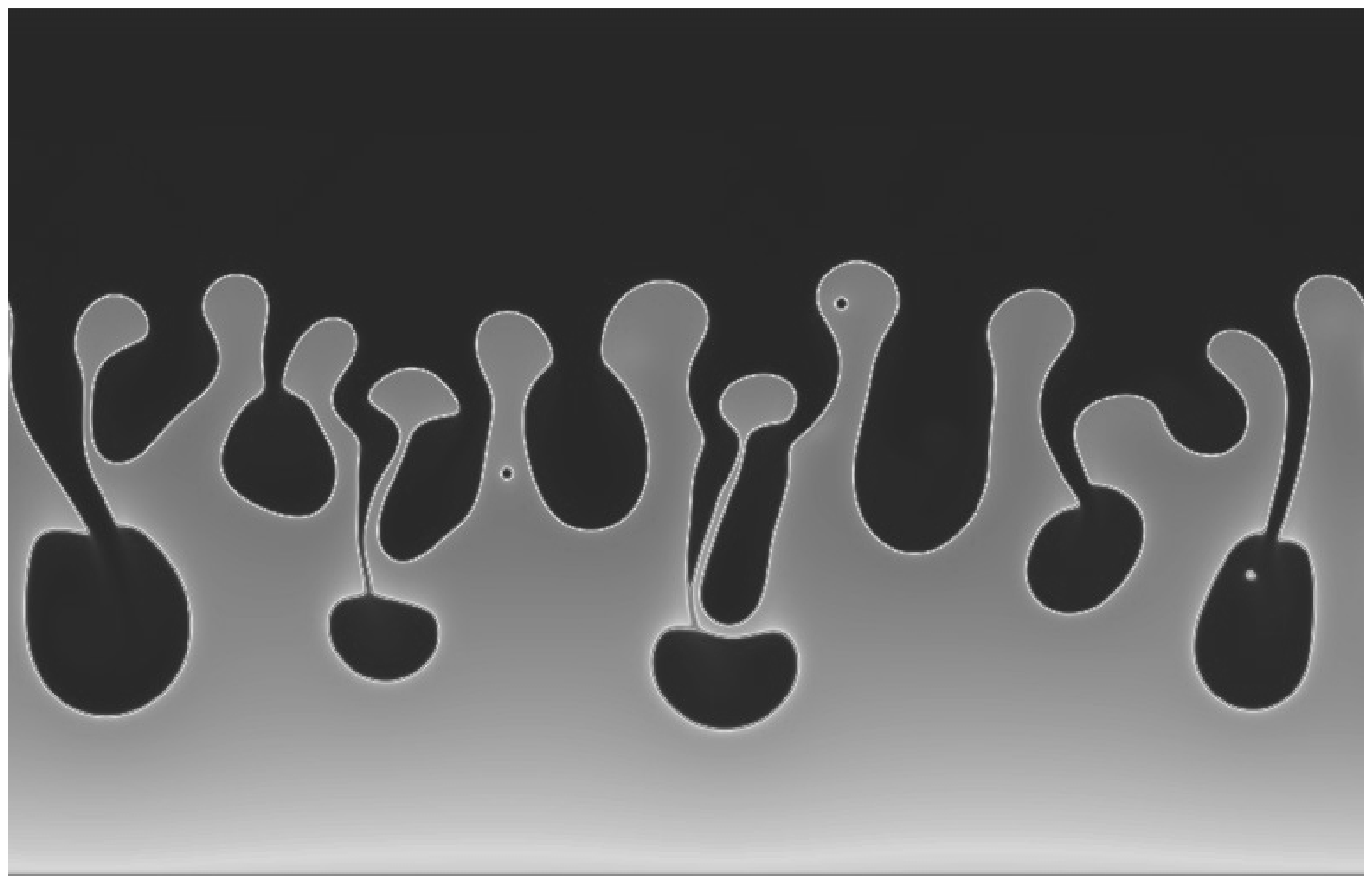}}
\resizebox{10cm}{!}
{\includegraphics{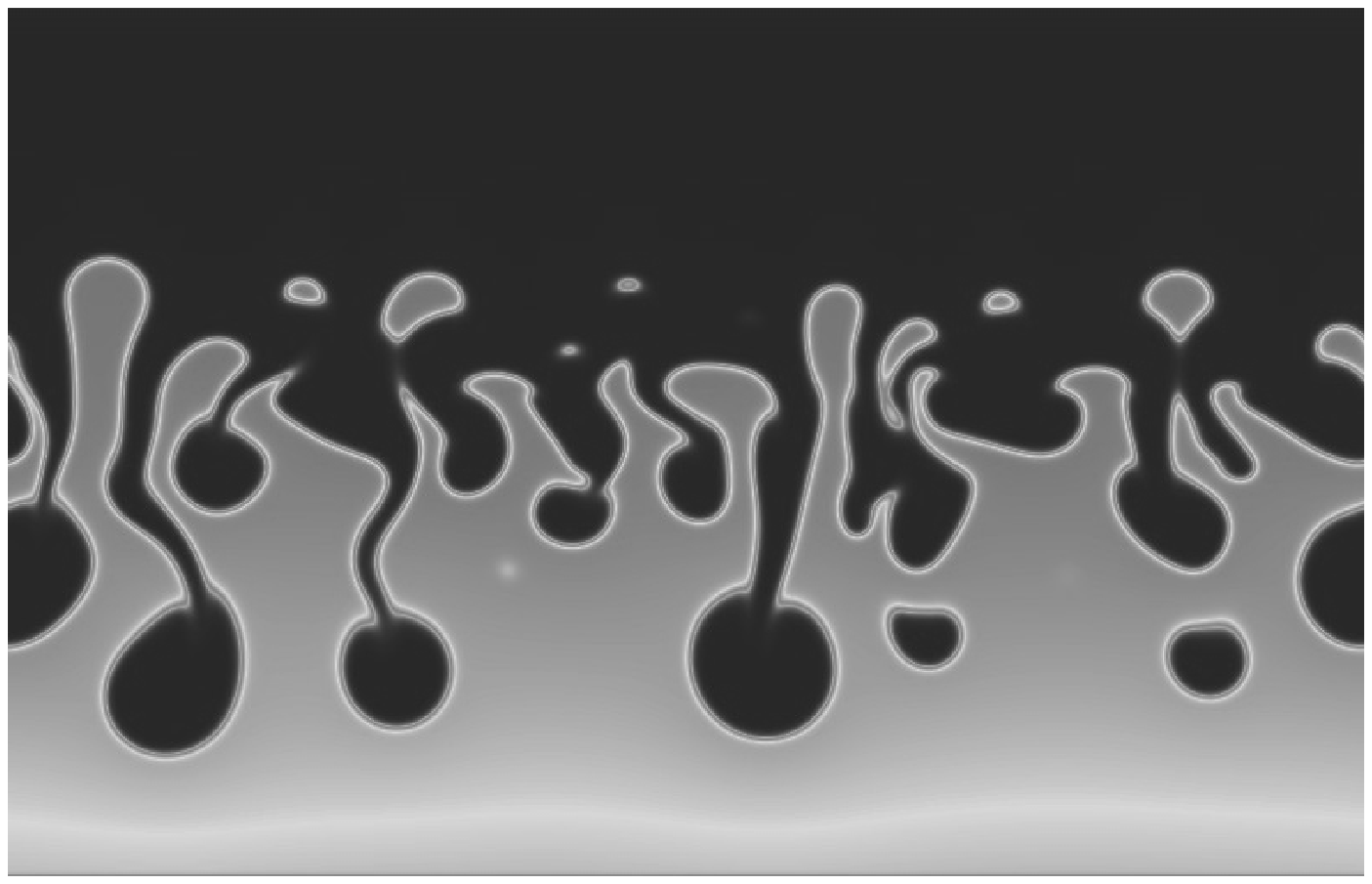}}
\caption{LB simulations of viscous fingering (after 
         $t=6000$ time steps) in miscible fluids, $G=0$ (top), 
         fluids with surface tension, $G=0.15$ (middle), 
         and miscible ($G=0$) reactive fluids (bottom).} 
\end{center}    
\end{figure} 

\begin{figure}[htbp] 
\rotatebox{-90} {   
\resizebox{10cm}{!}
{\includegraphics{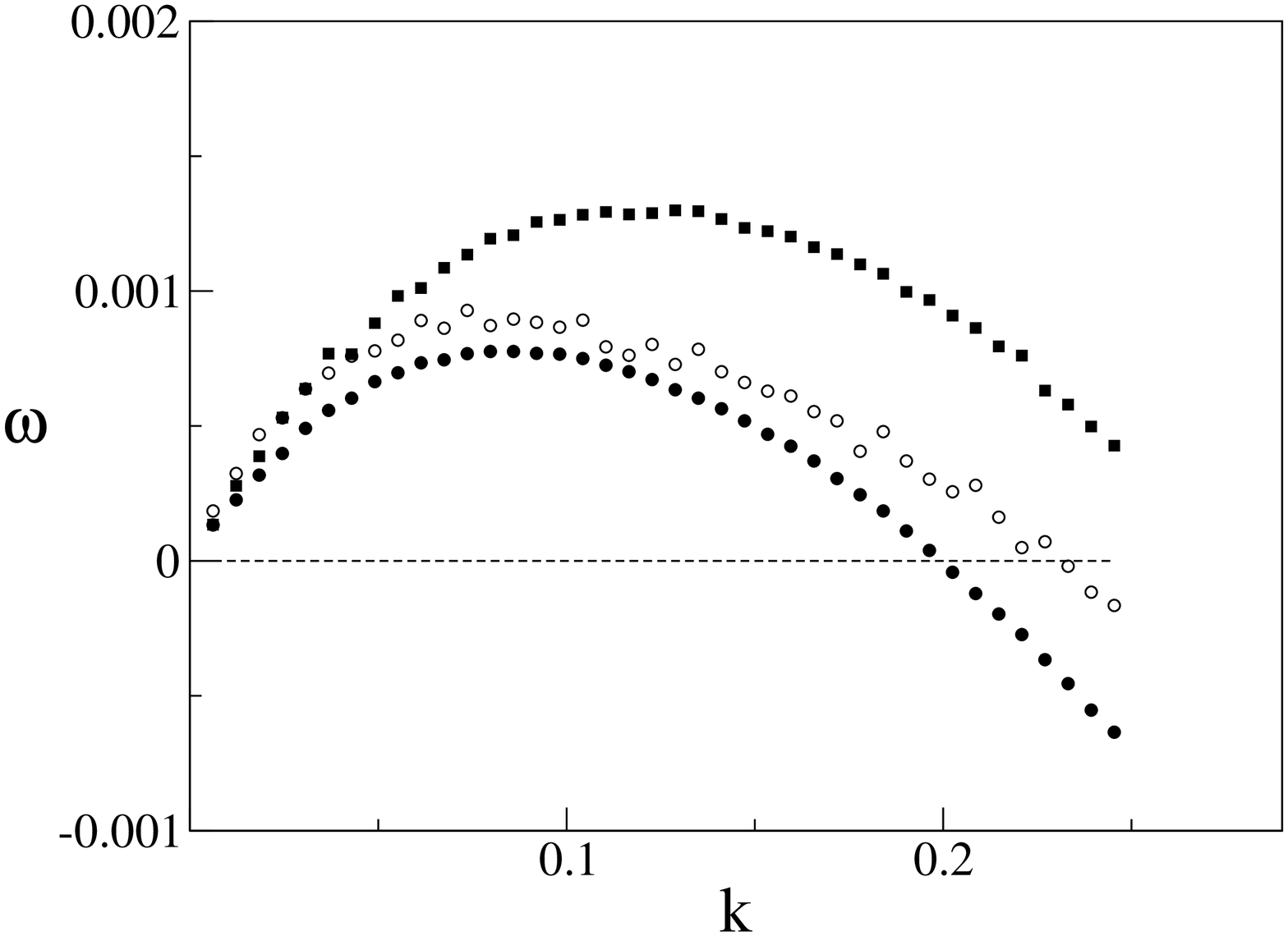}}
}
\caption{Dispersion curves for (a) miscible fluids ($G=0$; dots),
        (b) fluids with surface tension ($G= .05$; circles), and 
        (c) miscible reactive fluids ($\kappa=.001$; squares) 
        showing the effect of surface tension and of reactive 
        processes on  the most unstable wavelength and on 
        the cutoff wave number.}
\end{figure}

The purpose of this short paper is to demonstrate the applicability 
of the lattice Boltzmann method to the investigation of viscous 
fingering phenomena in spatially extended systems.
In particular we have shown how surface tension and reactive 
processes can modify the dynamics and the characteristics of the
fingering pattern. A quantitative analysis will be presented in
a forthcoming paper with a detailed description of the work.

\section*{Acknowledgments}

This work was supported by a grant from the 
{\em European Space Agency} and {\em PRODEX} (Belgium) 
under contract ESA/14556/00/NL/SFe(IC).



\begin{thebibliography}{0}

\bibitem{darcy}
See e.g. J. Bear, {\it Dynamics of Fluids in Porous Media}
(Dover, New York, 1988).

\bibitem{homsy}
G.M. Homsy, {\it Ann. Rev. Fluid Mech.} {\bf 19}, 271 (1987);
A. De Wit and G.M. Homsy, {\it J.~Chem. Phys.} {\bf 107}, 9609 
(1997); {\it ibid} {\bf 107}, 9619 (1997).

\bibitem{succi}
S. Succi, {\it The Lattice Boltzmann Equation for Fluid 
Dynamics and Beyond} (Clarendon Press, Oxford, 2001).

\bibitem{yoemans}
N. Rakotomalala, D. Salin, and P. Watzky, {\it J. Fluid Mech.}
{\bf 338}, 277 (1997);
K. Langaas and J.M. Yoemans, {\it Eur. Phys. J.} 
{\bf B15}, 133 (2000).

\bibitem{lutsko}
J.F. Lutsko, J.P. Boon, and J.A. Somers, in 
{\it Numerical Methods for the Simulation of Multi-Phase
and Complex Flow}, ed.  T.M.M. Verheggen
(Springer-Verlag, New York, 1992), p. 124.

\bibitem{hayot}
K. Balasubramanian, F. Hayot, and W.F. Saam, 
{\it Phys. Rev.} {\bf A36}, 2248 (1987);
N.S. Martys and H. Chen, {\it Phys. Rev.} 
{\bf E53}, 743 (1996).

\bibitem{doolen}
X. He, X. Shan, and G.D. Doolen,
{\it Phys. Rev.} {\bf E57}, R13 (1998).  

\bibitem{boon}
J.P. Boon, D. Dab, R. Kapral, and A. Lawniczak,
{\it Phys. Rep.} {\bf 273}, 55 (1996).

\bibitem{dewit}
A. De Wit and G.M.Homsy, 
{\it J.~Chem. Phys.} {\bf 107}, 8663 (1999);
A. De Wit, {\it Phys. Rev. Lett.} {\bf 87}, 054502-1 (2001).

\bibitem{qian}
Y. Qian, D. d'Humi\`eres, and P. Lallemand,
{\it Europhys. Lett.} {\bf 17}(6), 479 (1992).

\bibitem{experiment}
W.B. Zimmerman and G.M. Homsy, {\it Phys. Fluids}
{\bf A4}(9), 1901 (1992).

\bibitem{experim_reac}
M. B\"ockman and S.C. M\"uller, {\it Phys. Rev. Lett.}
{\bf 85}, 2506 (2000).


\end{thebibliography}
\end{document}